\documentclass[aps,reprint,twocolumn,superscriptaddress]{revtex4-1}

\usepackage{soul}
\setstcolor{red}
\usepackage{xcolor}
\usepackage{amsmath,amssymb,graphicx}
\usepackage{graphicx}
\usepackage{epstopdf}
\usepackage{mathtools}
\usepackage{upgreek}
\usepackage{notes2bib}
\usepackage{bm}
\usepackage{bm}
\usepackage{color}
\usepackage{braket}
\usepackage{textgreek}

\usepackage{float}

\usepackage{cancel}

\begin{document}

%\title{Coherent excitation of quasi-resonant optical modes}

\title{New perspective on chiral exceptional points with application to discrete photonics}

\author{ A. Hashemi}
 
\affiliation{Department of Physics, Michigan Technological University, Houghton, Michigan, 49931, USA}

\author{S. M. Rezaei}
\affiliation{Department of Physics, Michigan Technological University, Houghton, Michigan, 49931, USA}

\author{S. K. \"{O}zdemir}

\affiliation{Department of Engineering Science and Mechanics, and Materials Research Institute, The Pennsylvania State University, University Park, Pennsylvania 16802, USA}

\author{and R. El-Ganainy}
\email[Author to whom correspondence should be addressed: ]{ganainy@mtu.edu}
\affiliation{Department of Physics, Michigan Technological University, Houghton, Michigan, 49931, USA}
\affiliation{Henes Center for Quantum Phenomena, Michigan Technological University, Houghton, Michigan, 49931, USA}

\begin{abstract}
Chiral exceptional points (CEPs) have been shown to emerge in traveling wave resonators via asymmetric back scattering from two or more  nano-scatterers. Here, we provide a new perspective on the formation of CEPs based on coupled-oscillators model. Our approach provides an intuitive understanding for the modal coalescence that signals the emergence of CEPs, and emphasizes the role played by dissipation throughout this process. In doing so, our model also unveils an otherwise unexplored connection between CEPs and other types of exceptional points associated with parity-time symmetric photonic arrangements. In addition, our model also explains qualitative results observed in recent experimental work involving CEPs.  Importantly, the tight-binding nature of our approach allows us to extend the notion of CEP to discrete photonics setups that consist coupled resonator and waveguide arrays, thus opening new avenues for exploring the exotic features of CEPs in conjunction with other interesting physical effects such as nonlinearities and topological protections.

% {\color{red} OCIS code} 
  
\end{abstract}

%\ocis{(140.3490) Lasers, distributed-feedback; (060.2420) Fibers, polarization-maintaining; (060.3735) Fiber Bragg gratings; (060.2370) Fiber optics sensors.}
% REPLACE WITH CORRECT OCIS CODES FOR YOUR ARTICLE
% NOTE: \ocis{} IS ALIASED TO \pacs{} BUT MUST
% FORMAT THE TERMS CORRECTLY FOR EACH JOURNAL

\maketitle

\section{Introduction}
 
Recent works in non-Hermitian photonics have demonstrated a host of intriguing effects \cite{Musslimani2008,Makris2008,ElGanainy2008OB,Gao2015,Lu2017,Zhong2018winding,Zhong2018power,
Bliokh2019,Kuo2020,Arkhipov2020,Arkhipov2020quantum,Chen2020} and presented several opportunities for building new optical components and devices \cite{HosseinESH, Zhao2018THS, Zhang2018PLO, Hayenga2019, Zhong2019SES, ElGanainy2015SLA, Quesada2019, Qi2020PRApplied, Zhong2020PhysRevLett, Qi2021phy.rev.research}. A central concept in non-Hermitian physics is that of an exceptional point (EP) where two or more eigenvalues and the associated eigenstates of a non-Hermitian Hamiltonian coalesce and reduce the dimensionality of the eigenspace \cite{ElGanainy2018PRL, Feng2017JMP, Ozdemir2019JPL, Miri2019JPA}. To date, several implementations for exceptional points in photonics have been investigated  \cite{Gua2009PRL, ElGanainy2008OB, Peng2014NaturePhysics, Hodaei2014PTS, Wiersig2014ESF, Chen2017EPHS, Feng2014SML, Malzard2015PhysRevLett, Nada2017PhysRevB, Zhong2019SES, Zhong2020PhysRevLett}. A particular class of EPs that was recently explored theoretically and experimentally is the so called chiral EPs or CEPs, which are formed when the counter propagating modes of a traveling wave resonator coalesce into one traveling mode. In other words, at the CEP, the degenerate eigenmode has a preferred direction. While the concept of CEP  \cite{Ozdemir2010OE, Wiersig2011PRA,Wiersig2016SOEP,Wiersig2014ESF} with its potential applications \cite{Chen2017EPHS,Peng2016CM,Wiersig2020Photon.Res} has attracted considerable attention in recent years, its implementation has been so far confined to microring/microdisk resonators. Several new opportunities can be envisioned if this notion is extended to discrete photonic arrangements such as microcavity (or waveguide) arrays. These platforms offer unique optical features. For instance, they can serve as a testbed for exploring optical nonlinear effects \cite{Demetrious2003Nature}, spin-orbit coupling of light \cite{Carlon2019NatPhotonics}, topological protection \cite{Ozawa2019RevModPhys}, to mention just a few. However, in order to do so, an intuitive model that explains the physics of CEP is needed. \\

In this spirit, the purpose of this work is twofold: (1) to present an alternative and complementary approach for describing CEPs based on a coupled oscillators model and adiabatic elimination without resorting to full-wave scattering analysis; and (2) to utilize this approach to generalize and extend the notion of CEP to discrete photonic arrangements. A particular advantage of the method introduced here is that it emphasizes the role of dissipation in the emergence of CEPs whether it is a result of scattering or actual optical loss.  

\begin{figure*}[ht]
\centering
\includegraphics[scale=1]{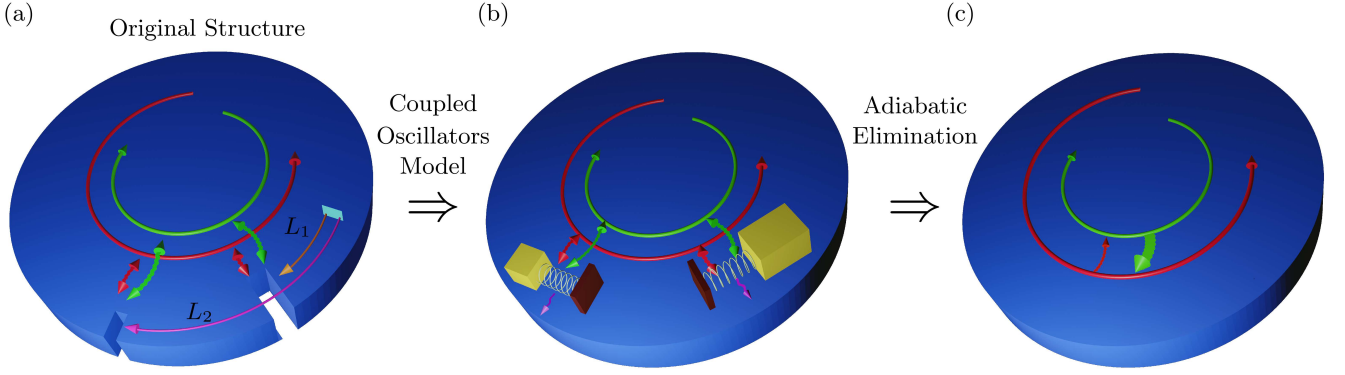}
\caption{(a) Schematic of a micro-disk resonator with two different nano-notches acting as Rayleigh scatterers to create a unidirectional coupling between clockwise(CW) and counterclockwise(CCW) modes. (b) A model for describing the system in (a) where the nano-scatterers are replaced by two harmonic oscillators that are far off-resonant with the optical modes of the resonator. (c) By adiabatically eliminating the oscillators, our mathematical model reduces to only interacting optical modes with asymmetric coupling coefficients.}
\label{Fig:notched_disk}
\end{figure*}

\section{Coupled oscillators model}
\textit{Model}---A typical implementation of a CEP involves the asymmetric back scattering between the clockwise (CW) and the counterclockwise (CCW) modes of a microring/microdisk resonators that can be tailored by using two different scatterers \cite{Ozdemir2010OE, Wiersig2011PRA}. Figure \ref{Fig:notched_disk} (a) shows a schematic depiction of this structure, where non-identical notches represent scattering centers. In order to engineer a CEP in this geometry, the two scatterers must introduce unidirectional coupling between the CW and CCW modes but without introducing a strong perturbation to their modal structures. Otherwise, these modes will lose their independent character (i.e. they cannot be treated as approximate modes of the system any more). The analysis of this problem in ref. \cite{Wiersig2016SOEP} consisted of two steps: (1) Modal degeneracy and Rayleigh scattering are used to show that the bidirectional coupling between the CW and CCW modes can be made asymmetric; (2) Full-wave simulations are employed to engineer the system's parameters to induce the unidirectional coupling. Here, we present a complementary but a more intuitive model for analyzing this problem. Our model highlights the important features of the system and the conditions needed to achieve a CEP. As we will see, this model will also allow us to generalize the notion of CEPs to discrete photonics configurations.\\

We start by recalling a well-known picture for light scattering from nanoparticles, namely that it can be viewed as a two-step process: (1) Light incident on a nanoparticle induces an electric dipole moment; (2) The oscillation of this dipole leads to the radiation of light. Within this picture, the microcavity-nanoparticles composite is described by the following linear coupled equations: $i d\vec{v}/dt=H\vec{v}$, where $\vec{v}=(a_{CW}, a_{CCW}, b_1, b_2)^T$. The first two components of $\vec{v}$ are the amplitudes of the CW and CCW optical modes while $b_{1,2}$ are the oscillation amplitudes associated with the nanoparticles, and the superscript $T$ denotes a matrix transpose. The effective Hamiltonian $H$ describing this model is given by:

\begin{equation}\label{Eq:Hamiltonian}
\hat{H}=
\begin{pmatrix}
	\omega_o & \epsilon & J_1e^{i\theta_1} & J_2e^{i\theta_2}\\
	\epsilon^* & \omega_o & J_1e^{-i\theta_1} & J_2e^{-i\theta_2}\\
	J_1e^{-i\theta_1} & J_1e^{i\theta_1} & \Omega_1 & 0\\
	J_2e^{-i\theta_2} & J_2e^{i\theta_2} & 0 & \Omega_2
\end{pmatrix}.
\end{equation}

\noindent Here, the real quantities $J_{1,2}$ are the strength of the coupling coefficients between the optical modes and the oscillators. Within the electric dipole approximation for the interaction between light and the nanoparticles, the values of these parameters depend on the strength of the electric field of the normalized optical modes at the location of the particles, which in principle can be obtained numerically by solving the microcavity problem in the absence of the particles. The phase factors $\theta_{1,2}$ result from the traveling wave nature of the modes and are given by $\theta_{1,2}=\beta L_{1,2}$, where $\beta$ is the propagation constant inside the ring waveguide and the distances $L_{1,2}$ are depicted in Fig. \ref{Fig:notched_disk} (a). In addition, the resonant frequency of optical modes is $\omega_o$ and those of the two oscillators are $\Omega_{1,2}$, respectively. In the Hamiltonian $\hat{H}$, we have also included direct coupling coefficients between the CW and CCW modes as described by $\epsilon$ in order to account for fabrication imperfections as we discuss later in detail. At the moment however, we assume $\epsilon=0$. Under this condition, the CW and CCW modes are not directly coupled but rather interact indirectly via the nano-scatterers. Also, note that there is no direct coupling between the two nano-scatterers. An important feature of this model is that losses due to optical absorption or radiation are treated on equal footing by taking the relevant frequency to be complex.  In order for the nano-scatterers to mediate the coupling between the CW and CCW waves without altering their modal structure,  the interaction between the optical modes and the scatterers must be non-resonant (i.e., Rayleigh scattering). \\
In other words, there must be a relatively large detuning between $\omega_o$ from one side and both $\Omega_{1,2}$ from the other side. Under this condition, one can perform adiabatic elimination \cite{Brion2007JPhysA} for the amplitudes associated with the nanoparticles $b_{1,2}$ and obtain an effective description for the amplitudes of the optical modes $a_{CW,CCW}$. By doing so (see Appendix \ref{Appendix:adiabatic_elimination} for details), we arrive at $i d\vec{a}/dt=\hat{M}\vec{a}$, where $\vec{a}=(a_{CW} , a_{CCW})$ and the reduced Hamiltonian $\hat{M}$ is given by:

\begin{equation}\label{Eq:Reduced_Hamiltonian}
\hat{M}=
\begin{pmatrix}
\omega_o+\alpha_1+\alpha_2 & \alpha_1 e^{2i\theta_1}+\alpha_2 e^{2i\theta_2} \\
\alpha_1 e^{-2i\theta_1}+\alpha_2 e^{-2i\theta_2} & \omega_o+\alpha_1+\alpha_2
\end{pmatrix} ,
\end{equation}
where, $\alpha_{1,2}=\frac{J_{1,2}^2}{\Delta_{1,2}}$ and $\Delta_{1,2}= \omega_o-\Omega_{1,2}$. The  procedure is valid in the limit when $\frac{J_{1,2}}{\Delta_{1,2}}\ll1$. By inspecting the above expression we find that $M_{11}=M_{22}$ is always satisfied. Additionally, it is easy to show that $2\big|\text{Im}[M_{11}]\big| \geq \big|M_{12}-M_{21}^*\big|$  as expected \cite{Wiersig2016SOEP} (see Appendix \ref{Appendix:inequality}). In the absence of dissipation, $\Omega_{1,2}$ are real and $M_{12}=M_{21}^*$, i.e. $M$ is Hermitian as expected. On the other hand, if the two oscillators exhibit different values of dissipation (either due to radiation or optical absorption) we find that the off diagonal elements are asymmetric. To illustrate this, let us assume that the oscillators $b_{1,2}$ have asymmetric dissipation, i.e. they are described by the complex detuning $\Delta_{1,2}=|\Delta_{1,2}|e^{-i\phi_{1,2}}$. It follows that the off-diagonal effective coupling elements are given by $M_{12,21}=|\alpha_1| e^{i(\pm 2\theta_1+\phi_1)} \left[1+\frac{|\alpha_2|}{|\alpha_1|} e^{i(\Delta \phi \pm 2\Delta \theta)}\right]$, where $\Delta \theta=\theta_2-\theta_1$ and $\Delta \phi=\phi_2-\phi_1$. The asymmetry of the off-diagonal elements $M_{12}$ and $M_{21}$ is thus introduced by the phase term $\Delta \phi \pm 2\Delta \theta$. The above expression provides an insight into the design restrictions needed to implement a CEP, i.e. to force only one of the off-diagonal elements to vanish and achieve unidirectional coupling between the CW and CCW modes (e.g., CW couples to CCW but CCW does not couple to CW mode or vice versa). We note that the system will implement a CEP regardless of the value of the unidirectional coupling strength as long as it is non-zero. From Eq.(\ref{Eq:Reduced_Hamiltonian}), we see that this condition can be achieved if $|\alpha_1|=|\alpha_2|=|\alpha|$, and either $\Delta \phi-2\Delta \theta=(2m+1)\pi$, $\Delta \phi+2\Delta \theta \neq (2l+1)\pi$; or  $\Delta \phi-2\Delta \theta \neq (2m+1)\pi$, $\Delta \phi+2\Delta \theta = (2l+1)\pi$ for any integers $m$ and $l$. Without any loss of generality, if we consider the first scenario, we find that the the maximum unidirectional coupling is obtained when $\Delta \phi-2\Delta \theta=(2m+1)\pi$ and $\Delta \phi+2\Delta \theta = 2l\pi$. In this case,  we find $M_{21}=0$, $M_{12}=2|\alpha|e^{i(2\theta_1+\phi_1)}$, and $M_{11,22}=\omega_o+|\alpha|e^{i\phi_1}\left[1+i(-1)^{l+m} \right]$. In other words, the introduction of unidirectional coupling between the CW/CCW modes inevitably causes dissipation, as expected. As a side remark, we note that for identical scatterers (i.e. $\alpha_1=\alpha_2$), both coefficients $M_{12}$ and $M_{21}$ can be zero simultaneously. This can be achieved by setting $\Delta \theta=(2m+1)\pi$. In other words, despite the presence of the two scatterers in this case, the two optical modes can be still decoupled

\textit{Relation to PT symmetry}---An important observation from the above results is that in order to operate at a CEP, the two oscillators (nanoparticles) must have asymmetric effective dissipation factors (due to absorption or scattering) which ensures that $\Delta \phi \neq 0$. This condition highlights the similarity between CEPs and EPs arising in PT-like systems which rely on inhomogeneous loss/gain distribution \cite{ElGanainy2007P}. To further elucidate on this connection, we use the numerical example cited above and rewrite $\hat{M}$ in the standing wave bases $\vec{a}_{sw}=(a_{+},a_{-})$ with $a_{\pm}=\frac{1}{\sqrt{2}}(a_{CW}\pm i a_{CCW})$. The new transformed Hamiltonian $\hat{M}_{sw}=\frac{1}{2}
\begin{pmatrix}
1 &i\\
1&-i
\end{pmatrix}
\hat{M}
\begin{pmatrix}
1&1\\
-i&i
\end{pmatrix}$ is then given by $\hat{M}_{sw}=\hat{M}_d+\hat{M}_{PT}$, with
$\hat{M}_d=(\omega_o+\alpha_1+\alpha_2) \hat{I}$, and $\hat{M}_{PT}=\alpha_1 e^{2i\theta_1} \hat{A}$; where  $\hat{A}=\begin{pmatrix} -i & i\\
-i& i
\end{pmatrix}$ and $\hat{I}$ being the $2 \times 2$ identity matrix. In this bases, the matrix $\hat{A}$ respects PT symmetry with the elements $\hat{A}_{11,22}$ representing effective loss and gain respectively, while the elements $\hat{A}_{12,21}$ denote Hermitian coupling between the modes $a_{\pm}$. Our model thus provides a unifying approach to describe both types of EPs and possibly open the door for engineering more versatile optical geometries that exhibit non-Hermitian singularities. This connection extends directly to systems respecting anti PT symmetry by recalling that if a Hamiltonian $\hat{M}_{PT}$ respects PT symmetry, then $i\hat{M}_{PT}$ is anti-PT symmetric.\\

\textit{Connection with experimental realizations}--- Next, We consider the case where $\epsilon \neq 0$. This situation arises in realistic experimental implementations due to defects or surface roughness that are inevitably introduced during fabrication. This introduces an additional coupling between the CW and CCW modes of the resonators. In recent experimental implementations of CEP, these undesired coupling effects were mitigated by using two fiber nanotips acting as nano-scatterers. The phase difference $\Delta \theta$ and the ratio between the dissipation factors associated with the two scatterers $\alpha_2$ and $\alpha_1$ were controlled by moving the nanotips along the ring perimeter or in the transverse direction (i.e. away or close to the ring surface). We now discuss this scenario within the context of our model and show that the latter can provide insight and account for these effects successfully. In this case, the reduced Hamiltonian after adiabatic elimination is modified according to $\hat{M} \rightarrow \hat{M}+\begin{pmatrix}
0 &\epsilon \\ \epsilon^* & 0
\end{pmatrix}$. In general, the degrees of freedom provided by the quantities $|\alpha_2/\alpha_1|$, $\Delta \phi$ and $\Delta \theta$ can allow for the cancellation of one of the coupling. As an illustrative example, consider the case where $\epsilon=|\epsilon| e^{-i(\phi_2-2\theta_2)}$. In this case, $\hat{M}_{21}=|\alpha_1| e^{i(- 2\theta_1+\phi_1)} \left[1+\frac{|\epsilon|+|\alpha_2|}{|\alpha_1|} e^{i(\Delta \phi - 2\Delta \theta)}\right]$, and the condition $\hat{M}_{21}=0$ is met when $\Delta \phi - 2\Delta \theta=(2m+1)\pi$ and $|\alpha_2|+|\epsilon|=|\alpha_1|$. Note that in that case $|\alpha_2| < |\alpha_1|$, which explains the tuning  of lateral position of the fiber nanotip in recent experimental works \cite{Chen2017EPHS}.

\section{CEP\lowercase{s} in photonic ring arrays}
\begin{figure}
\includegraphics[scale=1]{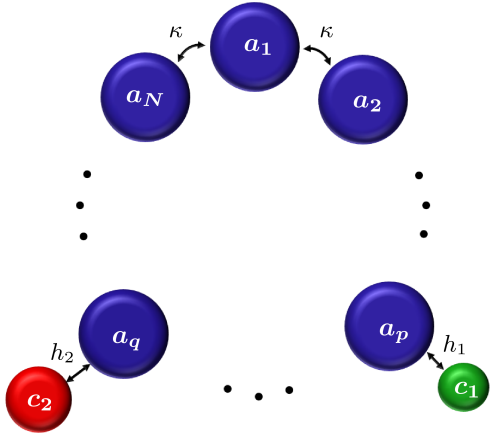}
\caption{Schematic of a discrete photonic array consisting of $N$, single mode microcavities (or waveguides) whose modal field amplitudes are denoted by $a_n$. Two perturbative cavities (shown in red and green) are introduced in the proximity of $p$th and $q$th elements of the array, and are characterized by field amplitudes with $c_1$ and $c_2$.}
\label{Fig:Array}
\end{figure}

The coupled oscillators approach presented in the previous section naturally lends itself to geometries other than microring resonators. In particular, it is very suited to explore the formation of CEPs in discrete photonic ring arrays without resorting to time-dependent perturbations  \cite{Longhi2014EPL}. These arrays can be formed by the evanescent coupling between waveguides or microcavities. While the structure of the evolution equations for both systems is identical (apart from replacing the propagation distance in  waveguides with time in microcavities), each offers a unique opportunity to investigate various physical effects \cite{Demetrious2003Nature, Carlon2019NatPhotonics, Ozawa2019RevModPhys}. Here, without any loss of generality, we will adopt the microcavity language. To this end, let us consider the case of $N$ coupled microcavities (Fig. \ref{Fig:Array}). We assume that each cavity supports only one optical mode at the resonant frequency of interest. Fig. \ref{Fig:Array} also shows two additional microcavities (red and green color) located at two different positions and termed as scatterers. The above system can be described with the following set of equations using the temporal coupled mode theory:

\begin{align}\label{Eq:ring_array}
\begin{split}
i\dot{a}_1 &=\omega_o a_1 +\kappa a_2 + \kappa a_{N} \\
i\dot{a}_n &=\omega_o a_n + \kappa a_{n+1} + \kappa a_{n-1}+\delta_{n,p} h_1 c_1+ \delta_{n,q} h_2 c_2 \\
i\dot{a}_N &=\omega_o a_N +\kappa a_1 + \kappa a_{N-1} \\
i\dot{c_1} &=\Omega_1 c_1 +h_1 a_p \\
i\dot{c_2} &=\Omega_2 c_2 +h_2 a_q \\
\end{split}
\end{align}

In Eq.\:(\ref{Eq:ring_array}), $a_n$ is the field amplitude of the resonant mode in cavity $n$, $c_{1,2}$ are the field amplitudes associated with the cavities representing the scatterers (see Fig.\:\ref{Fig:Array}), and the dot notation indicates time derivative. Also $\kappa$ is the nearest neighbor coupling constant along the ring structure and $h_{1,2}$ are the coupling coefficients between the scatterers and the corresponding cavities. Without loss of generality, in the absence of any gauge field, all the coupling coefficients can be taken to be real. In addition, $\omega_o$  is the eigenfrequency of the individual cavities forming the ring array, and  $\Omega_{1,2}$ represent the eigenfrequencies associated with the scatterers. Finally, $\delta_{x,y}$ is the usual Kronecker delta function. In the absence of the scatterers, the unperturbed ring array admits the solutions: $a_n(t)=e^{-i\mu_m t} e^{ik_m n}$, where $\mu_m=\omega_o+2\kappa \cos (k_m)$, subject to the quantization condition $k_m=\frac{2 m \pi}{N}$, $m=0,1,2,\ldots, N-1$. In the presence of perturbation, the above solution is not valid but nonetheless can be used as a new bases to re-express Eq.\:(\ref{Eq:ring_array}). Formally, this can be done via the symmetrized discrete Fourier transform $a_n (t)=\frac{1}{\sqrt{N}}\sum_{m=0}^{N-1} A_m(t) e^{ik_m n}$, and using the orthogonality relation $\sum_{n=0}^{N-1}e^{i(k_m-k_{m'})n}=N\delta_{m,m'}$, to obtain the following set of equations for $m=0,1,\ldots,N-1 $:

\begin{align}\label{Eq:discrete_Fourier}
\begin{split}
i\dot{A}_m &=\left[\omega_o + 2\kappa \cos(k_m) \right] A_m+ \frac{h_1}{\sqrt{N}} e^{-ik_m p} c_1\\
&+ \frac{h_2}{\sqrt{N}} e^{-ik_m q} c_2\\
i\dot{c_1} &=\Omega_1 c_1 + \frac{h_1}{\sqrt{N}} \sum_{l_1=0}^{N-1} e^{ik_{l_1}p} A_{l_1} \\
i\dot{c_2} &=\Omega_2 c_2 + \frac{h_2}{\sqrt{N}} \sum_{l_2=0}^{N-1} e^{ik_{l_2}q} A_{l_2}
\end{split}
\end{align}

By using adiabatic elimination of $c_{1,2}$ as before, we obtain:

\begin{align}\label{Eq:Reduced_ringarray}
\begin{split}
i\dot{A}_m &=\left[\omega_o + 2\kappa \cos(k_m) +\frac{h_1^2}{N \Delta_{1,m}}+\frac{h_2^2}{N \Delta_{2,m}} \right] A_m\\
&+\frac{h_1^2}{N} \sum_{l_1\neq m}^{} \frac{e^{i(k_{l_1}-k_m)p}}{\Delta_{1,l_1}} A_{l_1}\\
&+\frac{h_2^2}{N} \sum_{l_2 \neq m}^{} \frac{e^{i(k_{l_2}-k_m)q}}{\Delta_{2,l_2}} A_{l_2},
\end{split}
\end{align}

\noindent where $\Delta_{j,m}=\omega_o+2\kappa\cos(k_m)-\Omega_j$ for any $m=0,1,2,\ldots,N-1$ and $j=1,2$. Evidently, the two microcavities $c_{1,2}$ induce an interaction between all the supermodes. However, if the coupling terms $\frac{h_j^2}{N \Delta_{j,l}}$ are relatively small compared to the frequency detuning between the supermodes, one can neglect the off-resonant components and keep only the resonant interactions between counter propagating modes in the above equations, i.e. between modes characterized by the quantum numbers $m$ and $N-m$, excluding the mode $m=0$. In this case, Eq.\:(\ref{Eq:Reduced_ringarray}) simplifies to:

\begin{align}
\begin{split}
i\dot{A}_m &=\left[\omega_o + 2\kappa \cos(k_{m}) +\frac{h_1^2}{N \Delta_1}+\frac{h_2^2}{N \Delta_2} \right] A_m\\
&+\left(\frac{h_1^2}{N \Delta_1} e^{-2ik_mp} + \frac{h_2^2}{N \Delta_2} e^{-2ik_mq} \right)  A_{m'}\\
i\dot{A}_{m'} &=\left[\omega_o + 2\kappa \cos(k_{m}) + \frac{h_1^2}{N \Delta_1}+\frac{h_2^2}{N \Delta_2} \right] A_{m'}\\
&+\left(\frac{h_1^2}{N \Delta_1} e^{2ik_mp} + \frac{h_2^2}{N \Delta_2} e^{2ik_mq} \right)  A_m,
\end{split}
\end{align}

\noindent where we have used $\Delta_{j,m}=\Delta_{j,m'}=\Delta{j}$. If, as before, we assume that $\frac{h_{1,2}^2}{N\Delta_{1,2}}\equiv \chi e^{i\phi_{1,2}}$, we obtain:

\begin{align}\label{Eq:counter_propagating}
\begin{split}
i\dot{A}_m &=\left[\omega_o + 2\kappa \cos(k_{m}) +\frac{h_1^2}{N \Delta_1}+\frac{h_2^2}{N \Delta_2} \right] A_m\\
&+ \chi {e^{i(\phi_1-2k_mp)}}\left(1  +  e^{i[\Delta \phi -{2k_m} (q-p)]} \right)  A_{m'}\\
i\dot{A}_{m'} &=\left[\omega_o + 2\kappa \cos(k_{{m}}) +\frac{h_1^2}{N \Delta_1}+\frac{h_2^2}{N \Delta_2} \right] A_{m'}\\
&+ \chi {e^{i(\phi_1+2k_mp)}}\left(1  +  e^{i[\Delta \phi + {2k_m} (q-p)]} \right)  A_m.
\end{split}
\end{align}

From Eq.\:(\ref{Eq:counter_propagating}), it is clear that the effective coupling coefficients are asymmetric for $\Delta \phi \neq 0$. A CEP is formed when the value of one of these coefficients vanishes. This can be achieved for instance by satisfying the conditions $\Delta \phi -2k_m (q-p)= (2l+1) \pi$ and $\Delta \phi +2k_m (q-p) \neq (2l'+1) \pi$, for integers $l$, $l'$. Expressed explicitly, this translate into $\Delta \phi= \left(2l+1+\frac{4m(q-p)}{N} \right) \pi$ and $\frac{4m(q-p)}{N}\not\in\mathbb{Z}$. Moreover, for having the maximum unidirectional coupling $(\frac{4m(q-p)}{N}+\frac{1}{2})\in \mathbb{Z}$ should be satisfied.\\

%\begin{figure}
%\includegraphics[width=3.3in]{Fig_Pantagon.png}
%\end{figure}

\section{Conclusion}
 
In this work, we have presented a new approach to study CEPs. Our approach, which is based on coupled-oscillators model provides an intuitive and quantitative understanding for how CEPs emerge in traveling wave resonators, and emphasizes the role of dissipation. By using this model, we have demonstrated that the formation of CEPs in microring resonators and PT-symmetric EPs, in essence, share a common character, namely their formation rely on the same process of unbalancing the loss distribution. Finally, we have also shown how the concept of CEP can be extended to discrete photonics structures consisting of coupled resonator and waveguide arrays. Our results will open the door for investigating  the physics of CEPs in conjunction with other interesting physical effects such as nonlinearities and topological protections. Additionally, since the adiabatic elimination condition can be engineered in various physical platforms such as optomechanical and atomic setups, the simple intuition provided by our model may pave the way for implementing CEP in these systems.

\section{Acknowledgment}
R.E. acknowledges support from ARO (Grant No. W911NF-17-1-0481), NSF (Grant No. ECCS 1807552), and the Max Planck Institute for the Physics of Complex Systems.  S.K.O. acknowledges support from NSF (Grant No. ECCS 1807485), and AFOSR (Award no. FA9550-18-1-0235).

%\begin{acknowledgments}
%\end{acknowledgments}

\appendix
\section{The adiabatic elimination}\label{Appendix:adiabatic_elimination}
Adiabatic elimination is a standard approximation employed widely in physics. Yet, for a beginner who has never used this technique before, pedagogical references are lacking. For completeness, this section presents the major steps involved in this approximation. For simplicity, let us consider two identical oscillators coupled to a common oscillator (generalization to more complex configurations is straightforward):

\begin{align}\label{Eq:two_coupled_fields}
\begin{split}
&i\frac{da_1}{dt}=\omega_o a_1 + J_1b\\
&i\frac{da_2}{dt}=\omega_o a_2 + J_2b\\
&i\frac{db}{dt}=\Omega b +J_1^*a_1+J_2^*a_2,
\end{split}
\end{align}

\noindent where the detuning between resonant frequencies  $|\Delta\omega|\equiv |\omega_o-\Omega|$ is large compared to the magnitude of coupling coefficients $|J_{1,2}|$.  By using the exact transformation $a_{1,2}(t)\equiv\tilde{a}_{1,2}(t)e^{-i\omega_ot}$, $b(t)\equiv\tilde{b}(t)e^{-i\Omega t}$, Eq.\:(\ref{Eq:two_coupled_fields}) reduces to:

\begin{align}\label{Eq:tilde_coupled_fields}
\begin{split}
&i\frac{d\tilde{a}_1}{dt}=J_1e^{i\Delta\omega t}\tilde{b}\\
&i\frac{d\tilde{a}_2}{dt}=J_2e^{i\Delta\omega t}\tilde{b}\\
&i\frac{d\tilde{b}}{dt}=J_1^*e^{-i\Delta\omega t}\tilde{a}_1+J_2^*e^{-i\Delta\omega t}\tilde{a}_2.
\end{split}
\end{align}

The formal solution to the third line of Eq. (\ref{Eq:tilde_coupled_fields}) is:

\begin{equation}\label{Eq:Integral_equaiton}
\tilde{b}(t)=-i\int\limits_0^t\left[J^*_1\tilde{a}_1(\tau)+J^*_2\tilde{a}_2(\tau)\right]e^{-i\Delta\omega \tau}d\tau,
\end{equation}

\noindent where the initial condition $\tilde{b}(0)=0$ is assumed. By using integration by parts,  the above relation takes the form:

\begin{align}\label{Eq:Integral_eq_expand}
\tilde{b}(t)=&\frac{J^*_1}{\Delta\omega}\left[\tilde{a}_1(t)e^{-i\Delta \omega t}-\tilde{a}_1(0)-\int\limits_0^t\frac{d\tilde{a}_1}{d\tau}e^{-i\Delta\omega\tau}d\tau\right]\nonumber\\
+&\frac{J^*_2}{\Delta\omega}\left[\tilde{a}_2(t)e^{-i\Delta \omega t}-\tilde{a}_2(0)-\int\limits_0^t\frac{d\tilde{a}_2}{d\tau}e^{-i\Delta\omega\tau}d\tau\right].
\end{align}

Finally, by substituting this equation into the first line of (\ref{Eq:tilde_coupled_fields}), we obtain:

\begin{align}\label{Eq:reduced_a_tilde}
\begin{split}
&i\frac{d\tilde{a}_{n}}{dt}=\\
&\frac{|J_n|^2}{\Delta\omega}\left[\tilde{a}_n(t)-\tilde{a}_n(0)e^{i\Delta\omega t}-e^{i\Delta \omega t}\int\limits_0^t\frac{d\tilde{a}_n}{d\tau}e^{-i\Delta\omega\tau}d\tau\right]+\\
&\frac{J_n J_m^*}{\Delta\omega}\left[\tilde{a}_m(t)-\tilde{a}_m(0)e^{i\Delta\omega t}-e^{i\Delta \omega t}\int\limits_0^t\frac{d\tilde{a}_m}{d\tau}e^{-i\Delta\omega\tau}d\tau\right],
\end{split}
\end{align}

\noindent where $\{n,m\}=\{1,2\}$ or $\{n,m\}=\{2,1\}$.  It is worth noting that so far no approximation has been used in deriving the above results. In principle, Eq.\:(\ref{Eq:reduced_a_tilde}) can be solved iteratively by substituting the formal solution of  ${d\tilde{a}_{1,2}}/{d\tau}$ back into the same expression in a similar fashion to the procedure used in Dyson series \cite{Shankarbook}. This procedure will lead to perturbative expansion in terms of the parameter $|J_{1,2}|/\Delta \omega$. By invoking the condition $|J_{1,2}|\ll |\Delta\omega| \ll 1$ and retaining only the first term in the expansion integral, we arrive at:

\begin{align} \label{Eq:a_tilde_2}
&i\frac{d\tilde{a}_n}{dt}\simeq\nonumber\\
&\frac{J_n}{\Delta\omega}\left[J_n^*\tilde{a}_n(t)+J_m^*\tilde{a}_m(t)-\left(J_n^*\tilde{a}_n(0)+J_m^*\tilde{a}_m(0)\right)e^{i\Delta\omega t}\right].
\end{align}

Equation  (\ref{Eq:a_tilde_2}) describes an oscillator $\tilde{a}_n$ with resonance frequency $|J_n|^2/\Delta \omega$ that is coupled to an oscillator $\tilde{a}_m$, while at the same time is subject to   forcing term that oscillates at a frequency $\Delta \omega$. Since $|J_{1,2}|^2/\Delta \omega \ll \Delta \omega$, we can neglect the last term of the RHS of Eq.(\ref{Eq:a_tilde_2}) to obtain (this is valid only when the coupling coefficient is not large):

\begin{equation}
i\frac{da_n}{dt}\simeq \left(\omega_o +\frac{|J_n|^2}{\Delta\omega}\right)a_n(t)+\frac{J_n J_m^*}{\Delta\omega}a_m(t).
\end{equation}

The above analysis demonstrates that under the large detuning condition, the effect of the auxiliary oscillator $b$ is to shift the resonant frequency of oscillators $a_{1,2}$ in analogy to Lamb shift, and at the same time to induce an indirect coupling between them. The above steps can be generalized to any number of oscillators $a$ and $b$ by following exactly the same procedure outlined above.

\section{Constraints on matrix elements}\label{Appendix:inequality}
It is well known that for a passive non-Hermitian discrete Hamiltonian (i.e. without any gain elements), there exists a constraint on the corresponding matrix elements \cite{Wiersig2016SOEP}.  Here we confirm that elements of matrix $\hat{M}$ derived in Eq.\:(\ref{Eq:Reduced_Hamiltonian}) indeed satisfy this criterion, which reads: 
\begin{equation}
2\big|\text{Im}[M_{11}]\big| \geq \big|M_{12}-M_{21}^*\big|.
\label{Eq:inequality}
\end{equation}
Written explicitly, the left and right hand side of the above inequality are given by $2\big|\text{Im}[M_{11}]\big|=2\big| |\alpha_1|\sin \phi_1+|\alpha_2|\sin \phi_2\big|$ and $\big|M_{12}-M_{21}^*\big|=2\big| |\alpha_1|\sin\phi_1+e^{2i\Delta\theta}|\alpha_2|\sin\phi_2 \big|$, respectively.
By recalling that each scatterer introduces only optical dissipation, it follows that both $\sin\phi_1$ and $\sin\phi_2$ must be negative numbers. In other words, they have the same sign which automatically implies Eq.\:(\ref{Eq:inequality}) with the equality holding when $\Delta \theta=m \pi$ for an integer $m$.

%\section*{Data availability}
%The data that support the findings of this study are available from the corresponding author upon reasonable request.


\bibliography{Reference}

\end{document}